\documentclass[conference]{IEEEtran}
\IEEEoverridecommandlockouts
\usepackage{cite}
\usepackage{amsmath,amssymb,amsfonts}
\usepackage{algorithmic}
\usepackage{graphicx}
\usepackage{textcomp}
\usepackage{xcolor}
\def\BibTeX{{\rm B\kern-.05em{\sc i\kern-.025em b}\kern-.08em
    T\kern-.1667em\lower.7ex\hbox{E}\kern-.125emX}}

\usepackage{booktabs}
\usepackage{xurl}
\usepackage{makecell}
\usepackage{multirow} 
\usepackage[normalem]{ulem}
\usepackage[linesnumbered, ruled, vlined]{algorithm2e}
\usepackage{pifont}
\usepackage{amsfonts}
\usepackage{framed}
\usepackage{graphicx}
\usepackage{layout}
\newcommand{\bone}{\ding{182}}
\newcommand{\btwo}{\ding{183}}
\newcommand{\bthree}{\ding{184}}
\newcommand{\bfour}{\ding{185}}

\newcommand{\bsix}{\ding{187}}
\newcommand{\bseven}{\ding{188}}
\newcommand{\beight}{\ding{189}}
\newcommand{\bnine}{\ding{190}}

\newcommand{\blfootnote}[1]{%
  \begingroup
  \renewcommand\thefootnote{}\footnote{#1}%
  \addtocounter{footnote}{-1}%
  \endgroup
}

\begin{document}

\title{Exploring Memory Tiering Systems in the CXL Era via FPGA-based Emulation and Device-Side Management}

\author{
{Yiqi Chen\textsuperscript{*1}, Xiping Dong\textsuperscript{*1}, Zhe Zhou\textsuperscript{1,2}, Zhao Wang\textsuperscript{1,2}, Jie Zhang\textsuperscript{2},  Guangyu Sun\textsuperscript{$\dagger$}\textsuperscript{1}} \vspace{0.5em} \\
\textsuperscript{1}\emph{School of Integrated Circuits}, \textsuperscript{2}\emph{School of Computer Science}, 
\emph{Peking University}\\
\emph{\{yiqi.chen, zhou.zhe, wangzhao21, jiez, gsun\}@pku.edu.cn} \\  \emph{dxp@stu.pku.edu.cn}\\
}


\maketitle

\begin{abstract}
\label{sec:abstract}
The Compute Express Link (CXL) technology facilitates the extension of CPU memory through byte-addressable SerDes links and cascaded switches, creating complex heterogeneous memory systems where CPU access to various endpoints differs in latency and bandwidth. Effective tiered memory management is essential for optimizing system performance in such systems. 
However, designing an effective memory tiering system for CXL-extended heterogeneous memory faces challenges: 
1) Existing evaluation methods, such as NUMA-based emulation and full-system simulations like GEM5, are limited in assessing hardware-based tiered memory management solutions and handling real-world workloads at scale.
2) Previous memory tiering systems struggle to simultaneously achieve high resolution, low overhead, and high flexibility and compatibility.

In this study, we first introduce HeteroBox, a configurable emulation platform that leverages real CXL-enabled FPGAs to emulate the performance of various CXL memory architectures. HeteroBox allows one to configure a memory space with multiple regions, each exhibiting distinct CPU-access latency and bandwidth.  HeteroBox helps assess the performance of both software-managed and hardware-managed memory tiering systems with high efficiency and fidelity. Based on HeteroBox, we further propose HeteroMem, a hardware-managed memory tiering system that operates on the device side. HeteroMem creates an abstraction layer between the CPU and device memory, effectively monitoring data usage and migrating data to faster memory tiers, thus hiding device-side heterogeneity from the CPU. 
Evaluations with real-world applications show that HeteroMem delivers high performance while keeping heterogeneous memory management fully transparent to the CPU, achieving a 5.1\% to 16.2\% performance improvement over existing memory tiering solutions.

\end{abstract}

\section{Introduction}
\label{sec:introduction}

The increasing demand for DRAM in modern data centers has led to its significant share in both operational costs and power consumption, consuming approximately 37\% of the total cost and 33\% of the power, according to Meta~\cite{tpp_asplos23}. This dependency on DRAM is further entrenched by the DDR standard, which is designed around DRAM-specific access and maintenance commands, thereby limiting the integration of emerging memory technologies.

\blfootnote{* Co-first authors.}
\blfootnote{$\dagger$ Corresponding author.}

\begin{figure}[t]
  \centering
  \includegraphics[width=0.85\columnwidth]{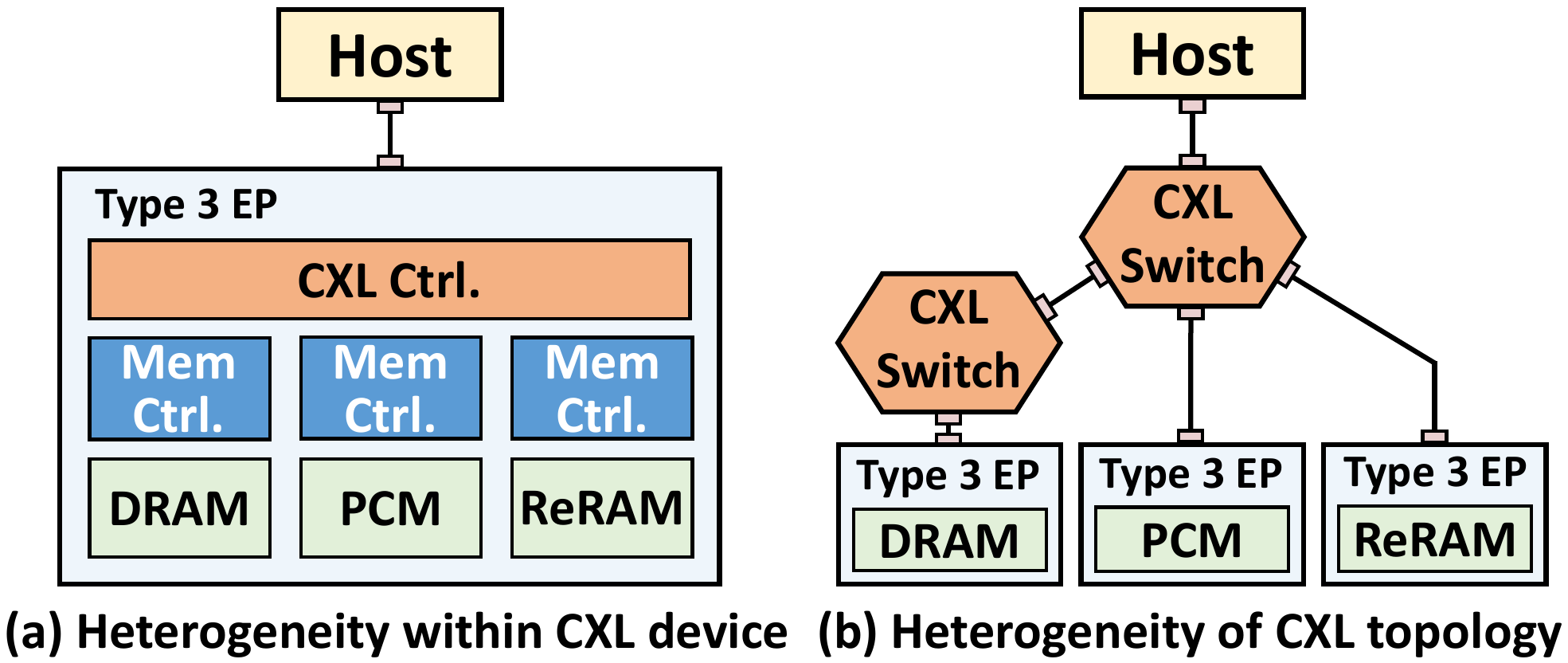}
  \vspace{-0.2cm}
  \caption{CXL-extended Heterogeneous Memory System}
  \vspace{-0.4cm}
  \label{fig:heterogeneous_system_overview}
\end{figure}

The emerging Compute Express Link (CXL) technique enables CPUs to access a variety of memory types by offering unified load-store semantics. This breakthrough paves the way for the adoption of alternative memory technologies, such as Managed DRAM~\cite{LeeJKKKKHKMLJKL19}, ReRAM~\cite{chen2020reram}, and 3D-XPoint/Optane~\cite{hady2017platform}, which promise larger capacities and lower costs~\cite{Subramanya_eurosys16}~\cite{Eisenman2018ReducingDF}~\cite{Kassa_atc21}, potentially reducing the Total Cost of Ownership (TCO) for data centers~\cite{sharma2024introductioncomputeexpresslink}. As shown in Figure~\ref{fig:heterogeneous_system_overview}-(a), different types of memory media can be integrated into single or multiple CXL memory devices. CPU can directly access the expanded memory via CXL channels, avoiding modifying the DDR memory controllers in existing CPUs. 
In addition to enhancing the diversity of memory media connected to the CPU, the CXL standard supports connecting the host to multiple endpoint devices via switches, enabling larger-scale memory expansion.
Additionally, CXL 3.0 introduces the capability for cascaded CXL switches~\cite{cxl}. As illustrated in Figure~\ref{fig:heterogeneous_system_overview}-(b), future CXL memory systems can feature a CXL switch connected to a host, another CXL switch, or a CXL memory device, thereby creating a complex topology.


The high flexibility of the CXL interconnect, while advantageous, also presents challenges in performance optimization. This duality is evident in several aspects. 
Firstly, new memory types (e.g., Optane) exhibit distinct latency and bandwidth characteristics. Integrating these  memory media into a system via CXL results in a high heterogeneity.
Secondly, the inclusion of CXL switches exacerbates such heterogeneity. 
As shown in Figure~\ref{fig:heterogeneous_system_overview}-(b), in a CXL-extended memory system with cascaded switches, different CXL memory devices may have varying numbers of CXL switches on the critical path to the host, resulting in different latency and bandwidth characteristics to the host CPU. In such systems featuring  heterogeneous memory types and topologies, we should strive to ensure that the CPU's memory access traffic takes the fast path rather than the slow one to preserve performance.

Prior research has endeavored to refine systems equipped with CXL memory from both system~\cite{tpp_asplos23, memtis_sosp23} and architectural perspectives~\cite{neomem, memstrata_osdi24}. The main idea of existing solutions is to leverage the access locality of workloads, identifying frequently accessed hot data and placing it into fast memory tiers. The remaining cold data is migrated to slow memory tiers to make room for hot data~\cite{mtm_ren2023, vtmm_eurosys23, simple_sc10}. 
However, prior works face challenges in optimizing the performance of the emerging CXL-extended heterogeneous memory systems. The reasons are as follows:

\noindent\emph{\textbf{Challenge 1}: Lack of Evaluation Platform} - 
Lacking real-world CXL-enabled CPUs and devices, previous studies rely on remote NUMA node emulation or software simulations to evaluate performance in CXL scenarios.
However, NUMA emulation-based solutions cannot arbitrarily configure the latency and bandwidth characteristics of memory and are challenging to apply to hardware-managed memory tiering systems. 
Meanwhile, software simulation approaches struggle with handling large workloads and assessing server CPU performance at a real scale due to low simulation speeds. 
Several studies test commercial CXL devices, but the latency and bandwidth attributes of these devices are not configurable, and CXL switches are currently unavailable.

\noindent\emph{\textbf{Challenge 2}: Low Performance of Software-Managed Memory Tiering System} - 
As detailed in Section~\ref{sec:background}, previous software-based memory tiering systems, which rely on various memory access profiling mechanisms including PTE scanning, hint-fault monitoring, and PMU sampling, struggle to accurately and efficiently identify hot and cold data due to hardware limitations. Meanwhile, the overhead of software-based data migration is significant.

\noindent\emph{\textbf{Challenge 3}: Poor Flexibility of Previous Hardware-Managed Memory Tiering System} - Previous hardware-based memory tiering systems, primarily proposed before the CXL era, achieve memory access profiling and data movement by incorporating custom hardware logic in the CPU. These intrusive modifications to the CPU prevent these solutions from adapting to various CXL-extended memory system configurations.

\noindent\textbf{Our Work:}
To address \textbf{Challenge 1}, we propose HeteroBox, an emulation platform based on a CXL-enabled FPGA. 
HeteroBox abstracts each CXL memory device as an addressable memory region that the host CPU can access with standard memory loads and stores. 
Each memory region can be assigned latency and bandwidth attributes corresponding to the emulated memory features.
For \textbf{Challenge 2}, we propose HeteroMem, a hardware-managed memory tiering system that efficiently profiles data hotness and migrates data to faster memory tiers using hardware logic. HeteroMem includes a profiling unit that tracks every memory request from the host CPU and identifies hot and cold pages, forwarding them to a migration unit. The migration unit is optimized for efficient data movement with high internal bandwidth.
For \textbf{Challenge 3}, we build HeteroMem logic on the device side, which acts as an abstraction layer between the CPU and device memory, hiding device-side heterogeneity from the CPU. 
A remapping unit in HeteroMem translates the address of memory request from host to device physical address, keeping migration completely transparent to CPU.
Our main contributions are as follows:

\begin{itemize}
    
    

    \item We propose HeteroBox, a configurable emulation platform leveraging real CXL-enabled FPGAs to emulate the performance of various CXL memory architectures. By carefully designing its architecture, we enable HeteroBox to support multiple emulated regions, each with distinct latency and bandwidth characteristics (Section~\ref{sec:heterobox}).

    \item We propose HeteroMem, a hardware-managed memory tiering system operating on the device side. We carefully design the architecture of HeteroMem to accurately and efficiently detect hot and cold data, and to manage data placement through effective data migration (Section~\ref{sec:heteromem}).

    \item We evaluate HeteroMem's performance on the HeteroBox emulation platform and explore various design choices for hardware-managed memory tiering systems. Our findings demonstrate that HeteroMem achieves a performance improvement of 5.7\% $\sim$ 17.6\% over several existing memory tiering systems, highlighting its effectiveness and potential as a superior memory management solution (Section~\ref{sec:evaluation}).


\end{itemize}

\section{Background \& Motivation}
\label{sec:background}

\begin{table*} \centering
    \caption{Comparison of HeteroMem with other Memory Tiering Systems}
    \vspace{-0.1cm}
    \label{tab:comparison_table}
    \resizebox{\textwidth}{!}{
    \Huge
        \begin{tabular}{c|*{3}{c}|*{3}{c}|*{2}{c}}
        \hline \noalign{\vspace{0.4ex}} \hline
                               & \multicolumn{3}{c}{Memory Access Profiling} 
                               & \multicolumn{3}{c}{Data Movement} 
                               & \multicolumn{2}{c}{Flexibility and Compatibility}  \\
                               & Mechanism & Overhead & Resolution 
                               & Mechanism & CPU Overhead & Bandwidth 
                               & CPU Transparent & Runtime Configurable  \\
        \midrule
        AutoNUMA~\cite{corbet2012autonuma}     & Hint-fault monitoring & High & Low & Moved by CPU & High & Low & No & Yes   \\
        TPP~\cite{tpp_asplos23}          & Hint-fault monitoring & High & Low & Moved by CPU & High & Low & No & Yes   \\
        DAMON~\cite{damon}     & PTE Scanning & High & Low & Moved by CPU & High & Low & No & Yes   \\
        MEMTIS~\cite{memtis_sosp23}       & PMU Sampling & Median & Median & Moved by CPU & High & Low & No & Yes   \\
        HeMem~\cite{hemem_sosp21}       & PMU Sampling & Median & Median & Moved by CPU & High & Low & No & Yes   \\
        MemPod~\cite{mempod_hpca17}       & Hardware Counters & Low & Median & Moved by Hardware & Low & High & Yes & No   \\
        Neomem~\cite{neomem}       & Hardware Hot Page Profiler & Low & High(only for hot pages) & Moved by CPU & High & Low & No & Yes   \\
        \midrule
        \textbf{HeteroMem}    & \textbf{Bidirectional Hardware Profiling} & \textbf{Low} & \textbf{High(both hot pages and cold pages)} & \textbf{Moved by Hardware} & \textbf{Low} & \textbf{High} & \textbf{Yes} & \textbf{Yes}   \\
        \hline \noalign{\vspace{0.4ex}} \hline
    \end{tabular}
    }
    \vspace{-0.4cm}
\end{table*}

\subsection{CXL-based Memory Expansion}
The Compute Express Link (CXL) protocol\cite{cxl}, based on the PCIe 5.0 physical layer, facilitates a cache-coherent, byte-addressable interconnect through high-efficiency SerDes links. This protocol composes three primary sub-protocols: \texttt{CXL.io}, \texttt{CXL.cache}, and \texttt{CXL.mem}. 
The \texttt{CXL.mem} sub-protocol is particularly noteworthy as it allows for direct CPU access to memory devices at cache-line granularity through the CXL interface. 
Unlike traditional DDR memory architectures with memory controllers in the host CPU, CXL incorporates memory controllers on the device side.
This decouples the architecture, allowing flexible integration of various memory technologies into CXL memory controllers while preserving the host CPU design. Such flexibility is crucial for meeting specific capacity, performance, and cost-effectiveness needs of modern memory systems~\cite{hynix_cxl_mem, samsung_cxl_mem, pond_asplos2023, cxl_enhanced_memory_function}.


CXL 2.0 introduces the CXL switch, which supports a single level of switching. The CXL switch allows one or multiple hosts to connect to one or more CXL devices. CXL 3.0 allows multiple switch levels, which enables more complex CXL memory device topologies. CXL 3.0 uses Port Based Routing (PBR) for inter-switch links to route CXL messages, allowing the software running on the host to perceive a flat topology ending at the leaf switch port. However, using a CXL switch introduces additional latency. According to~\cite{pond_asplos2023}, each CXL switch that a CXL request passes through adds about 70ns of latency. In a complex CXL memory system consisting of multiple switch levels, memory requests to different devices may pass through varying numbers of switches, resulting in a memory system with heterogeneous latency and bandwidth characteristics to the host CPU. 

However, due to the limited availability of real-world CXL-enabled CPUs and devices, the evaluation of CXL-extended memory systems faces significant constraints. Specifically, previous work often relies on remote NUMA node emulation~\cite{tpp_asplos23, memtis_sosp23} or software-based simulation methods~\cite{cxlanns_atc23} to assess the performance of CXL-based heterogeneous memory systems.
But real CXL-extended memory exhibits different latency and bandwidth characteristics compared to remote NUMA nodes according to~\cite{caption_micro23}. Furthermore, CXL decouples the CPU and memory interface, allowing for more complex modifications on the device side, which cannot be modeled by NUMA emulation-based platforms. 
Software-based simulation methods, however, struggle with handling large workloads and assessing server CPU performance at a real scale due to low simulation speeds. For instance, the simulation speed of gem5~\cite{lowepower2020gem5simulatorversion200} in full system mode is approximately 100 kilo instructions per second (KIPS), which is tens of thousands of times slower than emulation methods.  While some studies utilize CXL-enabled CPUs and devices~\cite{caption_micro23, cxlshm_sosp23, Sano_2023} for evaluation, their configurations are restricted by the available hardware  and can hardly adapt to other settings. 
These limitations highlight the need for an emulation platform that can efficiently and accurately evaluate the performance of CXL-extended heterogeneous memory systems.

\vspace{5pt}
\noindent\fbox{%
  \parbox{0.47\textwidth}{%
     \textbf{{Insight\#1:}}
     There's an urgent need for a practical, high-fidelity method to evaluate the complex CXL-extended heterogeneous memory system. 
  }%
}
\vspace{5pt}

\subsection{Memory Tiering System}



Many studies have focused on developing memory tiering systems to enhance the performance of heterogeneous memory systems. Specifically, we examine three critical aspects of memory tiering system design: (1) \emph{Memory Access Profiling}, (2) \emph{Data Movement}, and (3) \emph{Flexibility and Compatibility}. Table~\ref{tab:comparison_table} provides a summary comparison of prior work.

\begin{figure}[t]
  \centering
  \includegraphics[width=0.85\columnwidth]{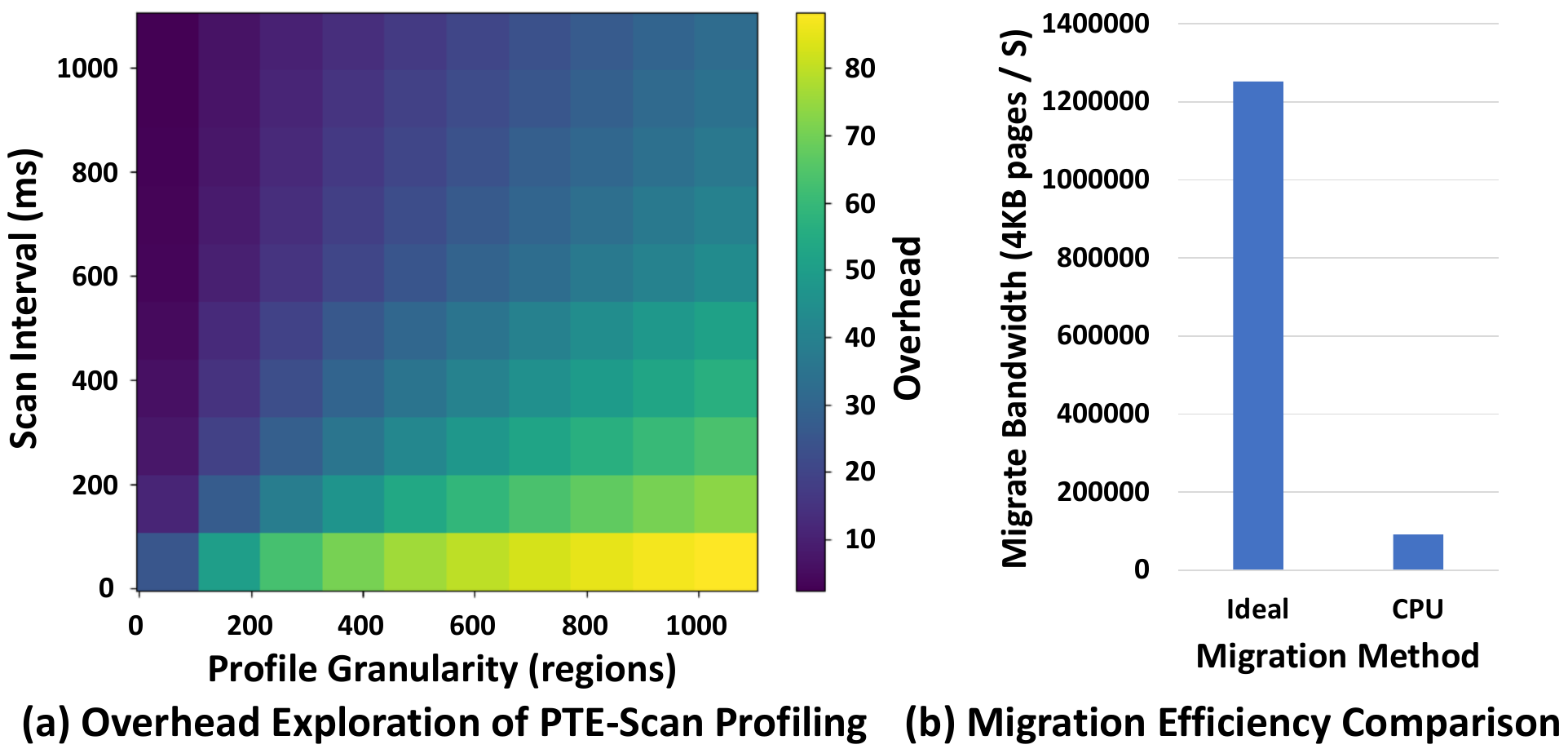}
  \vspace{-0.2cm}
  \caption{Overhead of Software Profiling and Data Movement}
  \vspace{-0.4cm}
  \label{fig:damon_overhead}
\end{figure}

\noindent{\textbf{Memory Access Profiling:}} 
The memory access profiling mechanism is crucial for managing heterogeneous memory systems and achieving performance improvements. 
However, compared to hardware-based methods, software-based memory access profiling methods suffer from high overhead and low resolution. Software-based methods primarily consist of three types: hint-fault monitoring, PTE scanning, and PMU-based sampling.
Hint-fault monitoring periodically poisons a portion of the memory pages, invalidating their PTEs and causing page faults when they are next accessed. TPP~\cite{tpp_asplos23} and AutoNUMA~\cite{corbet2012autonuma} utilize hint-fault monitoring to detect hot data in slow memory. PTE scanning uses the access bit in the PTE to detect hot data. When a page is accessed, the MMU automatically sets the access bit. A kernel thread periodically scans the memory space to check whether a page has been accessed during the last period. AMP~\cite{amp} uses PTE scanning to detect hot data in slow memory, while TMTS~\cite{tmts_asplos2023} uses PTE scanning to detect cold data in fast memory. PMU sampling leverages specific hardware counters in the CPU to profile hot data in slow memory. For example, Intel CPUs support using PEBS counters to record LLC miss events. Each time an LLC miss event occurs, the PEBS counter increments by one. When the counter reaches a user-defined threshold, it overflows and records the access address. HeMem~\cite{hemem_sosp21} and MEMTIS~\cite{memtis_sosp23} use PMU sampling to detect hot data in slow memory.

However, as discussed in previous work~\cite{neomem}, these mechanisms fail to achieve both low overhead and high resolution simultaneously.
As shown in Figure~\ref{fig:damon_overhead}-(a), we use DAMON~\cite{damon} to profile the CPU overhead introduced by PTE scanning as an example. DAMON is a data access monitoring framework subsystem for the Linux kernel, which can be configured to profile memory access with different spatial and temporal granularities. We vary the space granularity from 50 regions to 1050 regions and the time granularity from scanning the page table every 50ms to every 1050ms. We visualize the overhead as a heat map, where lighter colors indicate heavier overhead. We observe significant overhead when using small spatial or temporal granularities, with a maximum cost of 88.15\% CPU time. 

\vspace{5pt}
\noindent\fbox{%
  \parbox{0.47\textwidth}{%
     \textbf{{Insight\#2:}} Software-based memory access profiling methods suffer from high overhead and low resolution. 
  }%
}
\vspace{5pt}

\begin{figure*} [t]
    \centering
    \includegraphics[width=0.85\linewidth]{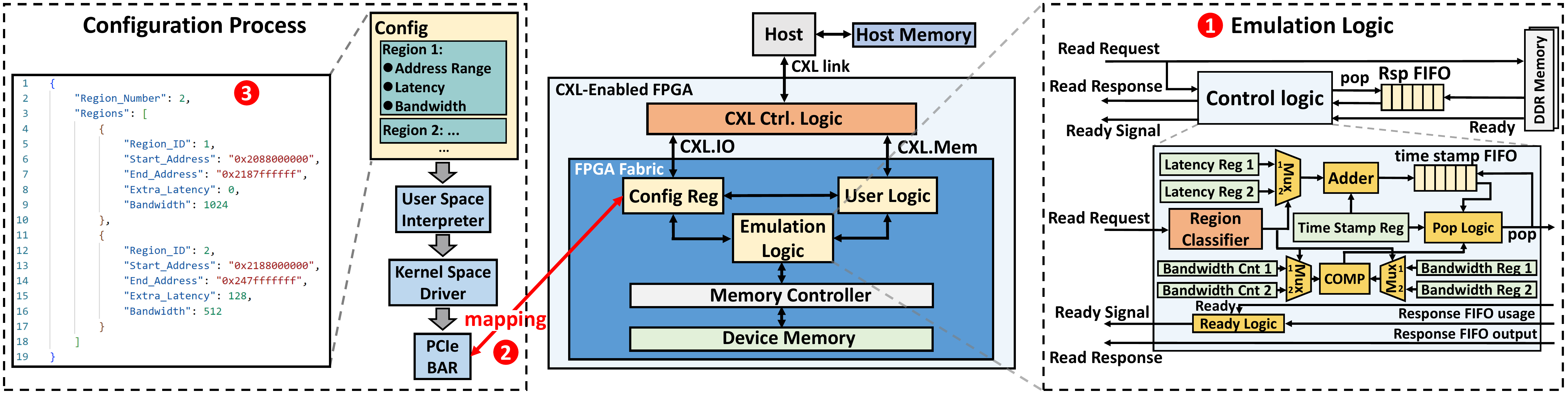} 
    \vspace{-0.2cm}
    \caption{Overview of HeteroBox Emulation Platform.}
     \label{fig:heterobox_overview}
     \vspace{-0.4cm}
\end{figure*}

\noindent{\textbf{Data Movement:}} 
After hot data is detected in slow memory, a data movement mechanism is required to migrate it to fast memory. Typically, the CPU can migrate data between fast and slow memory. 
Many previous works utilize the \texttt{migrate\_pages} API in the Linux kernel for this purpose. 
However, as demonstrated in Nimble~\cite{nimble_asplos19}, using the CPU for data migration incurs significant overhead and reduced bandwidth due to the software stack in the OS, including tasks like checking permissions and unmapping and remapping pages. Although some works attempt to mitigate these limitations by using transactional or parallel migration, they still rely on the CPU for data migration, which inevitably causes CPU overhead. Furthermore, using the CPU to migrate data between CXL memory devices requires reading the data out and then writing it back, which consumes the limited bandwidth of the CXL link.
As shown in Figure~\ref{fig:damon_overhead}-(b), we compare the bandwidth of using custom hardware logic, as described in Section~\ref{sec:heteromem}, to migrate data versus using the CPU to migrate data. For both methods, we use the 4KB page granularity and record the maximum number of pages that can be migrated per second. We observe that using custom hardware logic to migrate data achieves 12.9x higher bandwidth compared to using the CPU. 

\vspace{5pt}
\noindent\fbox{%
  \parbox{0.47\textwidth}{%
     \textbf{{Insight\#3:}} Using the CPU to move data results in higher overhead and lower bandwidth compared to device-side hardware-managed data movement.
  }%
}
\vspace{5pt}

\noindent{\textbf{Flexibility and Compatibility:}} 
Previous software-based CXL memory tiering systems can be classified into two categories.
One approach is to present CXL memory to the CPU as NUMA nodes, where each node consists of homogeneous memory. The advantage of using NUMA abstraction is that it can seamlessly leverage previously optimized mechanisms in the operating system.
Another approach is to use the DAX mode of CXL memory, in which the CXL memory device is exposed to the CPU as a file that supports load-store semantics. In this scenario, users must explicitly modify the source code of the application to use the DAX file as extended memory. 
Both using NUMA and DAX modes for heterogeneous memory management are not transparent to the CPU, reducing flexibility and compatibility.

Other works employ hardware-based memory management for better performance. For example, MemPod~\cite{mempod_hpca17} modifies the CPU-side memory controller to group different types of memory into several pods. The MemPod logic in the memory controller optimizes the performance of the heterogeneous memory system without CPU involvement. These hardware-based heterogeneous memory management systems remain transparent to CPU. However, due to architectural limitations, they require modifications to CPU hardware, making them unsuitable to adapt to various configurations of CXL memory.

A recent state-of-the-art work, NeoMem~\cite{neomem}, proposes integrating a hardware unit within the CXL memory device, which communicates with the host CPU to provide memory access information. However, NeoMem relies on CPU to process this information and perform subsequent data movement, making it non-transparent to CPU and reducing overall flexibility and compatibility.

\vspace{5pt}
\noindent\fbox{%
  \parbox{0.47\textwidth}{%
     \textbf{{Insight\#4:}} Previous software-based and hardware-based memory tiering systems fail to adapt to the complex configurations of CXL-extended memory systems.
  }%
}
\vspace{5pt}



To address these limitations, we first develop HeteroBox, an emulation platform designed to emulate a multi-tier CXL-extended heterogeneous memory system. Building on this platform, we introduce HeteroMem, a CXL-native, hardware-based memory tiering system. The core design philosophy of HeteroMem is to create an abstraction layer between host CPU and memory controller within the CXL device. This hardware-based abstraction layer is tasked with identifying hot data in slower memory and migrating it to faster memory. This entire process is completely transparent to host CPU. Further details are introduced in the following sections.


\section{HeteroBox Platform}
\label{sec:heterobox}


\subsection{Overview of HeteroBox Emulation Platform}

Figure~\ref{fig:heterobox_overview} illustrates the overview of HeteroBox emulation platform, where a CXL-enabled host processor (e.g., Intel 4$^{th}$-generation Xeon Sapphire Rapids CPU) connects to a CXL FPGA board that manages the device memory. 
We implement the emulation logic of HeteroBox between the memory controller and CXL controller, intercepting and processing all memory read requests. 
In the following description, we refer to the request address in the host address space as hPA and the address sent to the memory controller in the device as dPA. Typically, hPA equals dPA when no additional translation logic is involved.
The requests are assigned different latency and bandwidth characteristics according to their dPA.
Besides, we implement configuration registers for HeteroBox, mapping them to the PCIe BAR space to control the behavior of HeteroBox. A host-side software interface allows users to configure HeteroBox at runtime.
Meanwhile, users can implement custom hardware logic (e.g., hardware-based tiered memory management unit) on the HeteroBox emulation platform. 

Using a CXL-enabled CPU and FPGA, HeteroBox accurately emulates the performance characteristics of the CXL link, and its emulation logic replicates various latency and bandwidth characteristics. These capabilities allow HeteroBox to precisely model a wide range of CXL-extended heterogeneous memory systems.
In the subsequent part of this section, we will detail the design of HeteroBox's emulation logic and its configuration process.




\subsection{Emulation Logic Design}
The emulation logic of HeteroBox creates an illusion that the homogeneous DDR memory on the FPGA is divided into several heterogeneous parts, each with distinct latency and bandwidth characteristics. 
As shown in the right part of Figure \ref{fig:heterobox_overview} (\textcolor{red}{\bone}), the emulation logic contains a timestamp register, and for each abstract region, there is a latency register, a bandwidth configuration register and a bandwidth counter register. The timestamp register increments by one each cycle, while the latency registers and bandwidth configuration registers can be configured by users through \texttt{CXL.io} requests.
We focus on read requests since they are on the critical path of program execution.
When the ready signal is high, a read request can enter the emulation logic. HeteroBox first identifies the memory region of the dPA for the read request. It then attaches a timestamp, incremented by the latency register of the corresponding memory region, to the request as a tag. 
In practice, HeteroBox does not send the tag to the memory controller; instead, it pushes the tag into a FIFO and maintains the correspondence between each tag and the request's response.
When the read response is returned from memory, it is first pushed into a response FIFO. 
HeteroBox then checks the tag of the first element in response FIFO every cycle to determine if the value of the timestamp register exceeds the tag. If it does, HeteroBox pops the response out and returns it, adding a configurable latency to the region.

For bandwidth characteristics, HeteroBox maintains a bandwidth configuration register and a bandwidth counter register for each region. 
The bandwidth configuration register is user-configurable, while the bandwidth counter register increments by one whenever a response is returned from the corresponding region, and it resets at a user-defined interval. 
When the bandwidth counter register reaches the value set in the bandwidth configuration register for that region, HeteroBox blocks further responses from that region until the bandwidth counter register is reset in the next interval.

\subsection{Configuration Process}

As shown in the left part of Figure~\ref{fig:heterobox_overview} (\textcolor{red}{\btwo}), the configuration registers for HeteroBox are mapped to the PCIe BAR space of the host. These registers control the number of memory regions, the dPA range of each region, and their latency and bandwidth characteristics. The host CPU can read or write these registers through MMIO.

The HeteroBox platform supports multiple regions with distinct latency and bandwidth characteristics. To configure HeteroBox, we implement drivers in kernel space and an interpreter in user space. Users can configure HeteroBox using JSON files, as illustrated in the left part of Figure~\ref{fig:heterobox_overview} (\textcolor{red}{\bthree}). These files specify the number of memory regions, region IDs, start and end dPAs, latency (cycles at 200MHz), and bandwidth (bandwidth configuration register). The JSON file is processed by user space interpreter, which sends the information to the kernel space driver. The driver then writes to PCIe BAR space to update configuration registers in HeteroBox.


\section{HeteroMem Solution}
\label{sec:heteromem}

\subsection{Overview of HeteroMem System}
\begin{figure}[t]
  \centering
  \includegraphics[width=0.85\columnwidth]{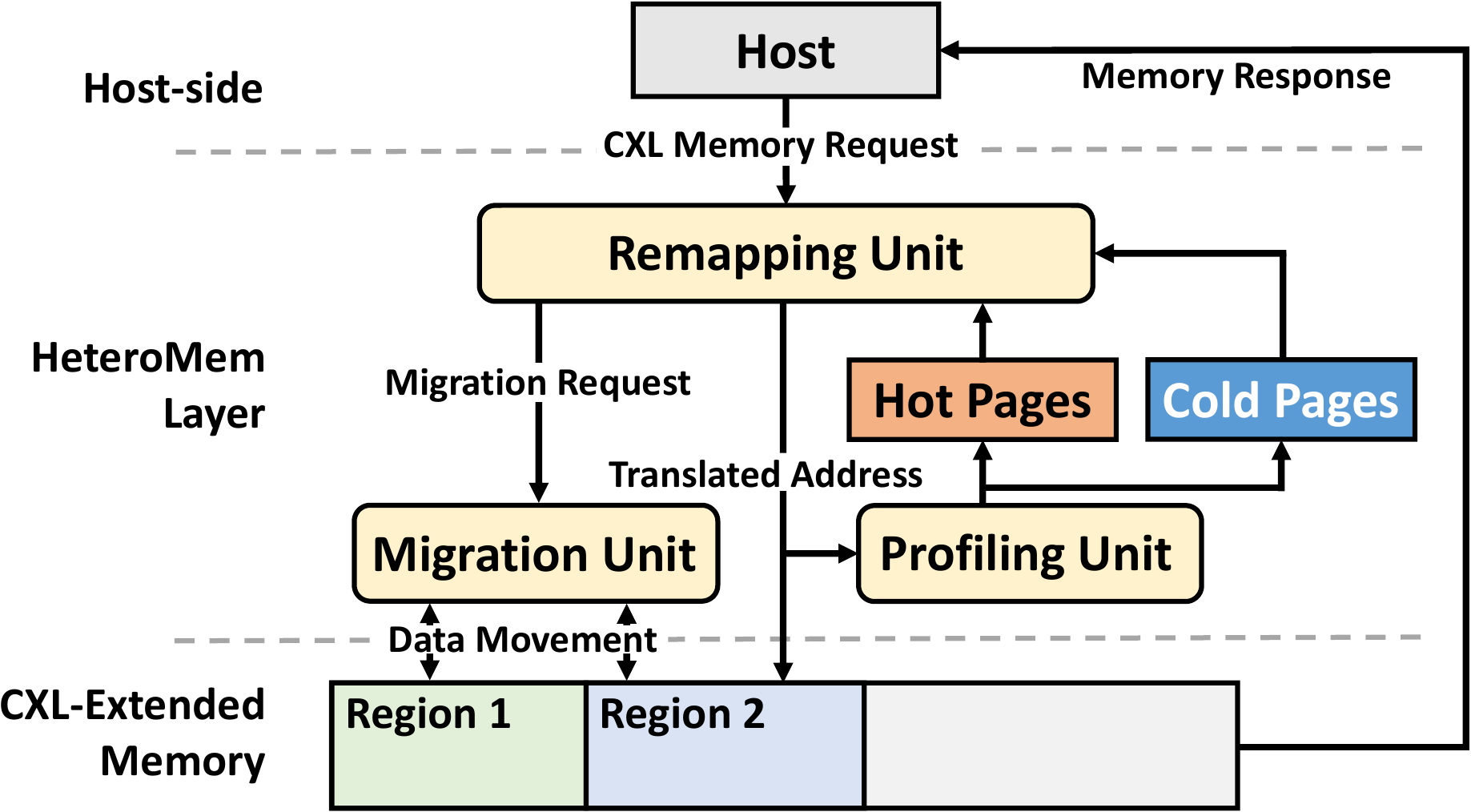}
  \vspace{-0.2cm}
  \caption{System Overview of HeteroMem}
  \vspace{-0.4cm}
  \label{fig:heterobox_system_overview}
\end{figure}


The HeteroMem memory tiering system acts as an intermediate layer between host CPU and CXL-extended memory. 
As shown in Figure~\ref{fig:heterobox_system_overview}, the HeteroMem system comprises three main components: the \textbf{Remapping Unit}, the \textbf{Profiling Unit}, and the \textbf{Migration Unit}. 
Additionally, we provide software interface support for HeteroMem memory tiering system, including a software driver in the kernel and BAR registers in the device for configuration and profiling. 
For the following evaluation of HeteroMem memory tiering system, we use a two-tier memory system, consisting of a fast memory tier and a slow memory tier, without losing generality.


\begin{figure*}[t!]
    \centering
    \includegraphics[width=1.0\linewidth]{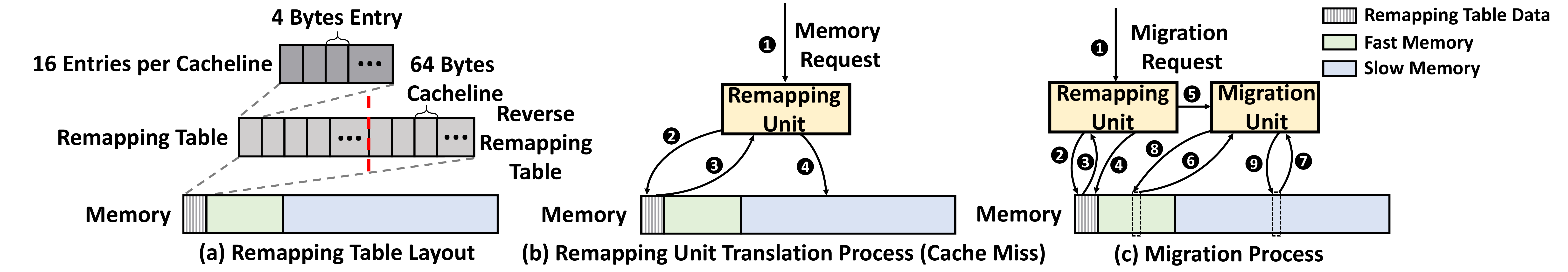}
    \vspace{-0.5cm}
    \caption{Remapping Unit Working Process.}
    \label{fig:remap_table_working_process}
    \vspace*{-0.4cm}
\end{figure*}
\begin{figure}[t]
  \centering
  \includegraphics[width=\columnwidth]{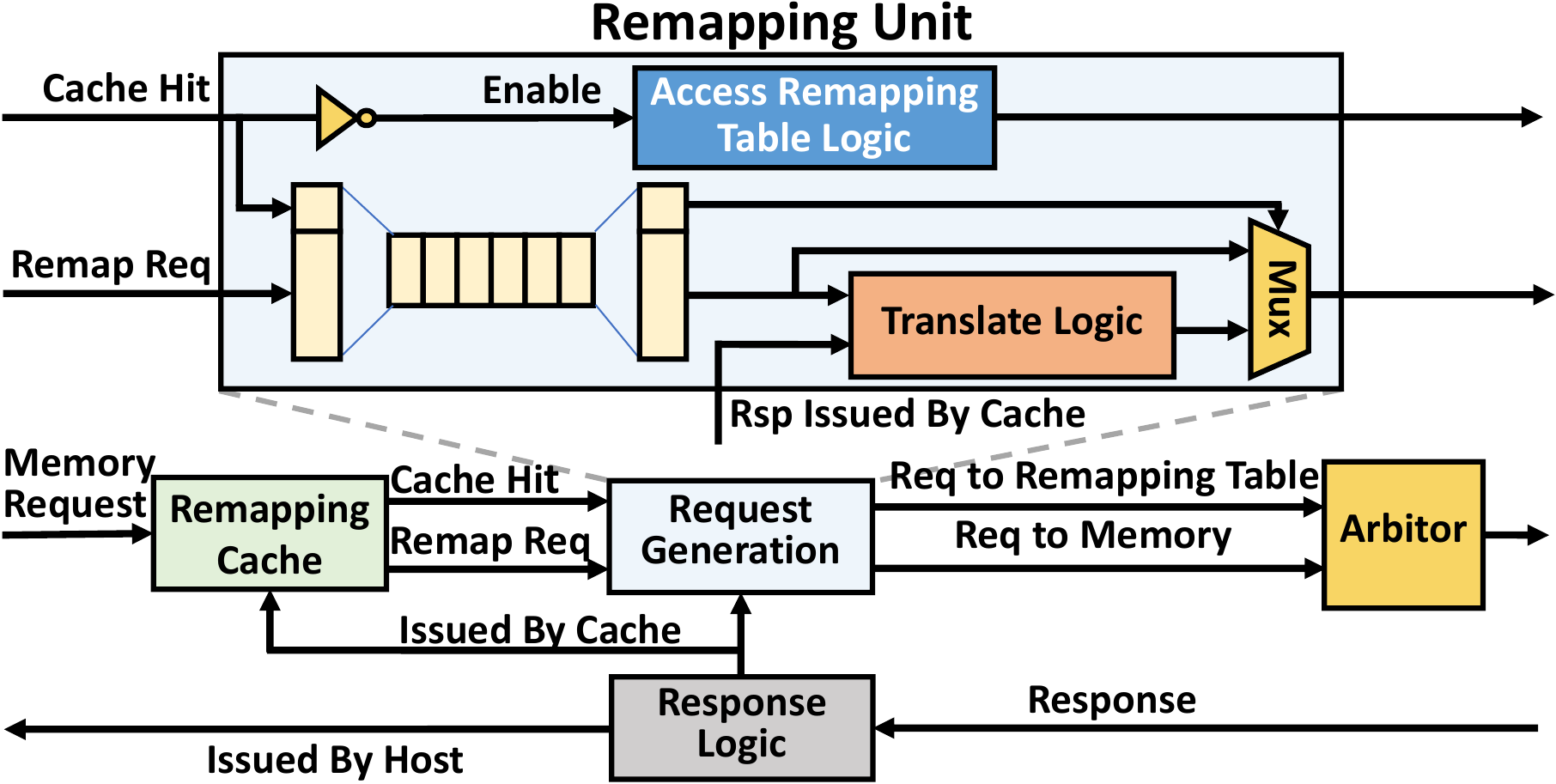}
  \vspace{-0.5cm}
  \caption{Block Diagram of the Remapping Unit}
  \vspace{-0.4cm}
  \label{fig:remap_table}
\end{figure}

\noindent{\textbf{Remapping Unit:}} 
HeteroMem manages heterogeneous memory system entirely on the device side, migrating hot data to fast memory and cold data to slow memory. To keep this migration transparent to CPU, the device needs a translation layer to map hPA to dPA. We implement a Remapping Unit to handle this, translating memory requests and forwarding them to subsequent modules. Since the Remapping Unit is on the critical path of memory requests, it must minimize added latency to ensure optimal system performance.

\noindent{\textbf{Profiling Unit:}} 
The Profiling Unit assesses the hotness and coldness of data in CXL-extended memory using translated memory requests from the Remapping Unit. It records access patterns, profiling hot data in slow memory and cold data in fast memory, temporarily buffering the dPA of detected cold data. When hot data is identified, the Profiling Unit signals Remapping Unit to initiate a migration event, providing the dPA for hot data and the dPA of previously buffered cold data for subsequent migration transaction.

\noindent{\textbf{Migration Unit:}} 
The Migration Unit initiates data migration upon receiving requests from the Remapping Unit, which contain dPAs for hot and cold data identified by the Profiling Unit, and swaps their memory locations. To keep migrations transparent to the CPU, the Remapping Unit must be updated whenever a migration occurs. 


\subsection{Remapping Unit}

The Remapping Unit provides an additional translation layer between the CPU and the device. It needs to be carefully designed to introduce minimal overhead and keep the translation process transparent to the CPU.

\noindent{\textbf{Memory Layout:}} 
As shown in Figure \ref{fig:remap_table_working_process}-(a), the entire memory space consists of a fast memory region and a slow memory region. 
The Remapping Unit stores the metadata for address translation at the beginning of the fast memory, which is software-reserved to avoid access by the host CPU. 
When the machine powers on, the Remapping Unit initializes this metadata by sending a series of write requests to the reserved fast memory.
This metadata includes a remapping table array and a reverse remapping table array, both consisting of 4-byte entries, each containing a page index (for dPA). 
HeteroMem manages memory in 4KB page granularity, so a 4-byte page index can describe a memory space of up to 16TB. For larger memory spaces, a longer index (e.g., 8 bytes) can be used.
Accessing the remapping table array with a page index yields the translated page index, while the reverse remapping table contains the inverse mapping of the remapping table.

\noindent{\textbf{Architecture of Remapping Unit:}} 
The Remapping Unit's architecture, shown in Figure \ref{fig:remap_table}, works as follows: when a memory request arrives, it first checks the remapping cache, which stores metadata for translating requests. If the required metadata is found, the request is translated and sent to memory. On a cache miss, the request is buffered in a FIFO, and the remapping cache issues a memory read to retrieve the needed remapping table data. Once the data returns, the cache is updated, and the buffered request is translated and forwarded to memory.
Figure~\ref{fig:remap_table_working_process}-(b) illustrates the translation process of the Remapping Unit when a remapping cache miss occurs.

\begin{figure*} [t]
    \centering
    \includegraphics[width=0.85\linewidth]{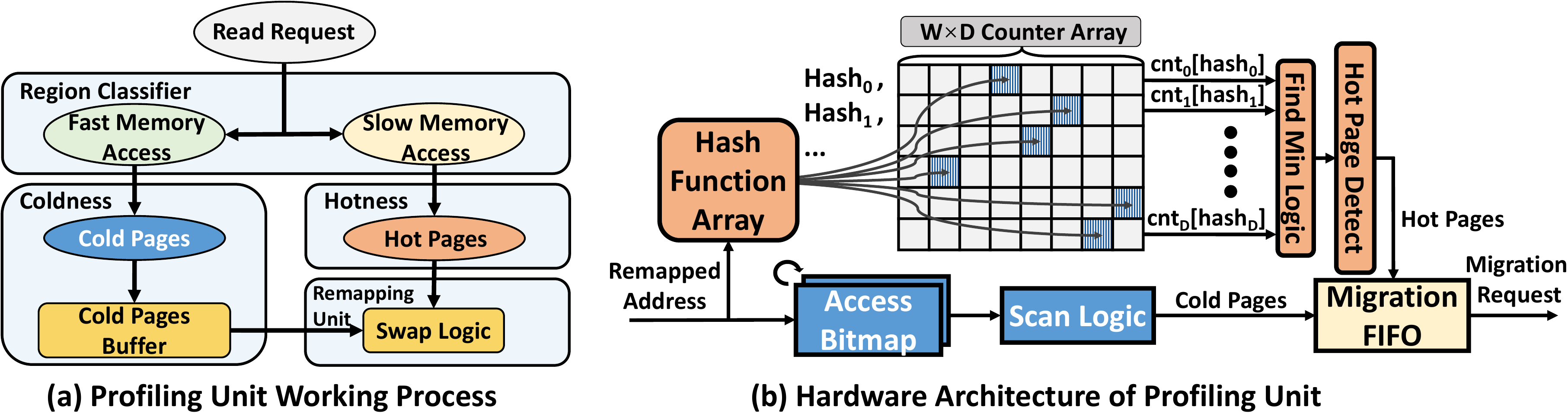} 
    \vspace{-0.2cm}
    \caption{Profiling Unit Overview.}
     \label{fig:profiling_unit_overview}
     \vspace{-0.4cm}
\end{figure*}

\noindent{\textbf{Migration Transaction:}} 
As shown in Figure~\ref{fig:remap_table_working_process}-(c), when the Remapping Unit receives a migration request from the Profiling Unit (\bone), it initiates a migration transaction. The migration request contains a pair of page indexes located in fast memory and slow memory, respectively. The page indexes in the migration request have already been translated. 
Upon receiving the migration request, the Remapping Unit blocks subsequent requests from the host to ensure the atomicity of the migration. 
Then, the Remapping Unit issues two memory read requests to the reverse remapping table in the memory to retrieve the hPAs of the hot and cold page (\btwo). 
When the read responses containing the hPAs return (\bthree), the Remapping Unit updates the remapping table data and the reverse remapping table data according to the two hPAs and two dPAs (\bfour). 
The Remapping Unit concurrently sends an enable signal to the Migration Unit after issuing the memory read request to overlap the memory read latency. 
The Migration Unit then reads the data of the two pages at the two page indexes specified by the migration request (\bsix \bseven), swaps their indexes, and writes them back (\beight \bnine). 
After the Migration Unit has completed the migration process and the Remapping Unit has updated its remapping and reverse remapping table data, subsequent memory requests are allowed to proceed.

\subsection{Profiling Unit}


The Profiling Unit is responsible for profiling the hotness and coldness of data in memory. In our HeteroMem design, we profile data hotness and coldness at a 4kB page granularity. 
The Profiling Unit is linked after the Remapping Unit, off the critical path, to capture the dPA trace. 
We profile only read requests, as write requests are generally not on the critical path of program execution.
As shown in Figure~\ref{fig:profiling_unit_overview}-(a), in the case of a CXL-extended memory system with two tiers, the Profiling Unit splits the read requests into two streams: one targeting fast memory and the other targeting slow memory. 
For the read request stream targeting fast memory, the Profiling Unit detects cold pages based on their access history. These detected cold pages are temporarily buffered in a cold pages buffer. Since cold pages tend to remain cold for a relatively long time~\cite{tmts_asplos2023}, they typically remain cold when fetched from the buffer in subsequent processes.
For the read request stream targeting slow memory, the Profiling Unit measures the hotness of each page based on its access frequency, reporting a page as hot if its hotness exceeds a given threshold. Upon detecting a hot page, the Profiling Unit fetches a cold page from the cold pages buffer and sends the detected hot page and the fetched cold page to the Remapping Unit.
The Remapping Unit then manages the data movement, swapping pages between fast memory and slow memory. The number of these hot and cold page pairs sent to the Remapping Unit is limited by a given constraint (e.g., 32 pairs of pages per 100,000 cycles).


\noindent{\textbf{Hotness Profiling:}} 
For memory read requests targeting fast memory, we use a Count-Min Sketch~\cite{count_min_sketch} structure to profile the hotness of each page. Count-Min Sketch is a hash-based algorithm that can estimate heavy hitters within a stream. As shown in Figure~\ref{fig:profiling_unit_overview}-(b), the main body of the Count-Min Sketch is a $\textbf{D}*\textbf{W}$ counter array, where $\textbf{D}$ is the depth of the array and $\textbf{W}$ is the width. The $\textbf{W}$ entries in the same row are referred to as a lane. When a request comes in, its dPA is mapped to $\textbf{D}$ different lanes using $\textbf{D}$ different hash functions. The mapped counters are incremented by one, and when the counters overflow, they remain at their maximum value. 
The minimum value of the $\textbf{D}$ mapped counters is then sent to the subsequent hot page detection logic. If this minimum value exceeds a given threshold, the accessed page is marked as hot.
Each counter has an associated hot bit that is set when the corresponding page is detected as hot. A page will not be reported as hot if all of its $\textbf{D}$ hot bits are already set, thus avoiding repeated reporting of the same hot page. All counters are reset periodically. Since the counters in the Count-Min Sketch provide an approximate value of access frequency and the error bound of the Count-Min Sketch is determined by the length of the stream, we control the reset period to limit the sketch's error bound.

\noindent{\textbf{Coldness Profiling:}} 
As detailed in Section~\ref{sec:heteromem}.D, the Migration Unit uses a swap mechanism to move hot data to fast memory. 
Therefore, when a hot page is detected in slow memory, we need to find a cold page in fast memory to swap its location with the detected hot page. We use a ping-pong bitmap to record the access history of each page in fast memory. The ping-pong bitmap structure consists of two bitmap arrays, each entry being a bit corresponding to a page in fast memory.
In a given period, one bitmap, say bitmap $A$, is used to record memory access in the current period. When Profiling Unit receives a memory request, it sets the bit corresponding to the accessed page in bitmap $A$. Meanwhile, Profiling Unit scans the other bitmap, say bitmap $B$, to identify pages with their corresponding bit in bitmap $B$ unset as cold pages. In the next period, we reset all bits in bitmap $B$ and switch the functions of bitmap $A$ and bitmap $B$. 
This approach ensures the detected cold pages were not accessed during the last period, thus accurately identifying cold pages.


\subsection{Migration Unit}
The Migration Unit moves data between fast memory and slow memory. 
On the device side, we lack knowledge about whether an address contains valid data, so every migration operation is executed as a swap operation, swapping the data between fast memory and slow memory. 
The Migration Unit takes in a pair of dPAs and an enable signal. 
When the enable signal is set high, the Migration Unit stages the two input dPAs, reads the data from these dPAs into a buffer, and then writes the data to the swapped dPAs to complete the process. 
The Migration Unit issues read requests to the memory controller in a non-blocking manner and initiates write requests as soon as a read response returns, while simultaneously blocking incoming memory requests from the host during migration, thereby minimizing idle cycles and maximizing migration bandwidth.

\label{sec:heterobox_software_interface}




\section{Implementation}
\label{sec:implementation}

\begin{table}[t]
\caption{Evaluation System Configuration}
\vspace{-0.1cm}
\label{table:system_config}
\resizebox{0.485\textwidth}{!}{

\begin{tabular}{|l|l|}
\hline
Host CPU   & Single socket Intel$^\circledR$ Xeon 6430 CPU @ 2.10GHz  \\ 
                        \cline{2-2} 
                      & \begin{tabular}[c]{@{}l@{}}32 Cores, hyperthreading disabled. \\ 
                        60MB Shared LLC\end{tabular} \\ 
                        \hline
DDR Memory            & 32GB DDR5 4800MHz x 2         \\ 
                        \hline
CXL Memory & One Intel$^\circledR$ Agilex$^{\texttt{TM}}$ I-Series FPGA Dev Kit @400 MHz \\ 
                        \cline{2-2} 
                      & \begin{tabular}[c]{@{}l@{}}Hard CXL 1.1 IP on PCIe Gen5 x16 \\ 
                        16GB 2-Channel DDR4-2666 DRAM\end{tabular}      \\ 
                        \hline
Software Environment  & Linux kernel version v6.3 \\
                        \hline

\end{tabular}
}
\vspace{-0.4cm}
\end{table}
\begin{figure}[t]
  \centering
  \includegraphics[width=0.85\columnwidth]{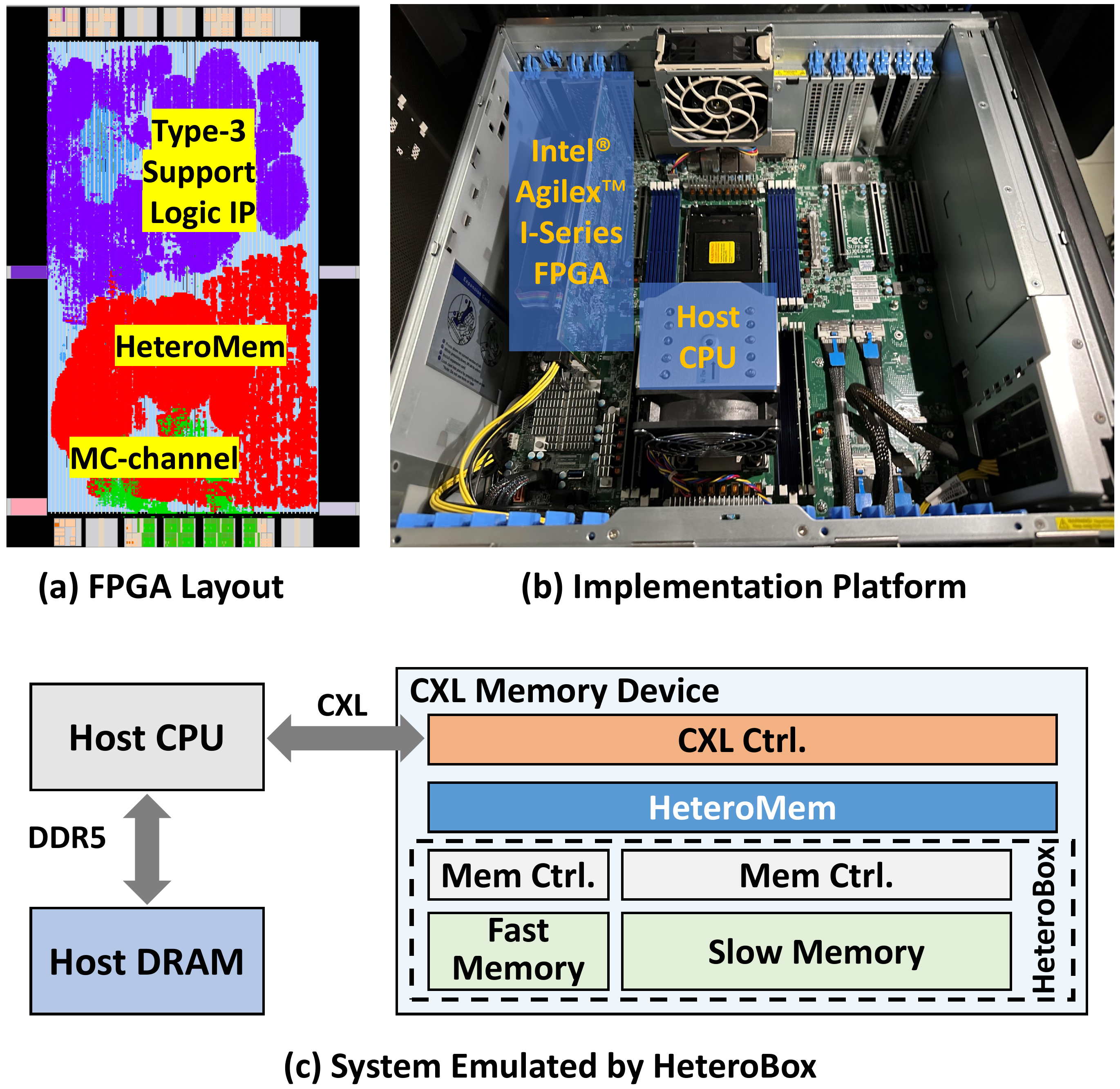}
  \vspace{-0.2cm}
  \caption{FPGA-based Prototyping System}
  \vspace{-0.4cm}
  \label{fig:implementation}
\end{figure}

We build our HeteroBox emulation platform on a real CXL system, as detailed in Table \ref{table:system_config}. The system setup includes a single-socket Intel$^\circledR$ Sapphire-Rapids$^\texttt{TM}$ CPU paired with a CXL-enabled Intel$^\circledR$ Agilex$^\texttt{TM}$-7 I-Series FPGA that acts as CXL memory (CXL 1.1, Type-3 device). The FPGA is equipped with dual-channel DDR4-2666 memory with a capacity of 16GB, while the host CPU is equipped with 32GB $\times$ 2 dual-channel DDR5-4800 memory.

We implement HeteroMem based on the Linux kernel v6.3. The CXL memory on the FPGA is exposed to the software as a CPU-less NUMA node. To compare the performance of HeteroMem with other baseline memory tiering systems, we modified the NUMA nodes enumeration code and split the CXL memory NUMA node into two fake NUMA nodes, corresponding to the fast memory and slow memory emulated by HeteroBox.
We use Intel$^\circledR$ FPGA CXL IP (Memory Expander Type 3) and add about 7000 lines of Verilog RTL custom logic code to build HeteroBox and HeteroMem, along with approximately 1000 lines of C++ simulation environment code and about 1000 lines of modifications in the Linux kernel for HeteroBox and HeteroMem driver and fake NUMA nodes logic. 
The HeteroBox is publicly available in the provided repository\footnote{~\url{https://github.com/memory-of-star/HeteroBox}}.

\begin{figure*} [t]
    \centering
    \includegraphics[width=0.85\linewidth]{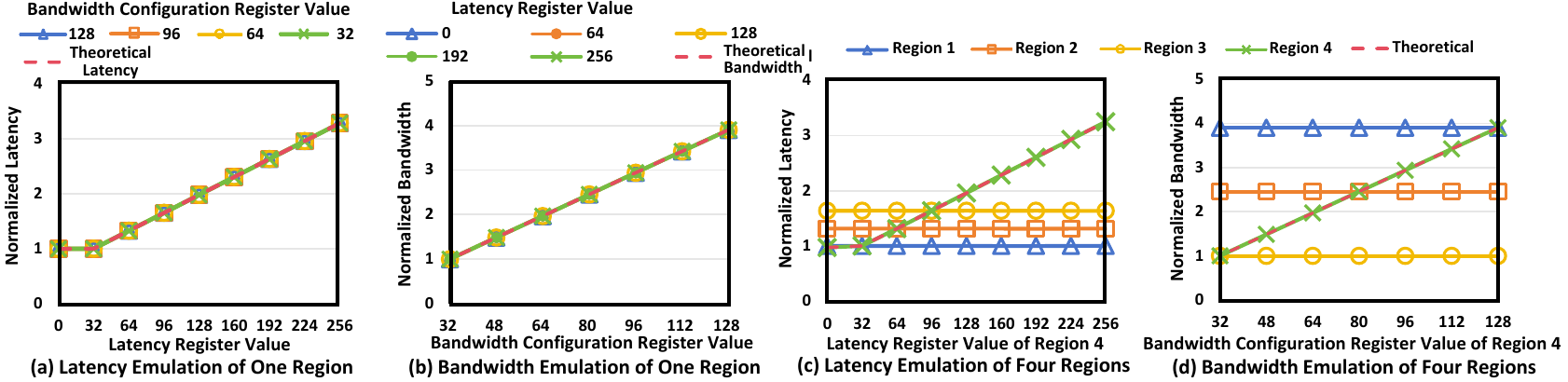}
    \vspace{-0.2cm}
    \caption{Validation of HeteroBox Emulation Platform}
     \label{fig:mlc_bandwidth_latency}
     \vspace{-0.1cm}
\end{figure*}
\begin{figure*} [t]
    \centering
    \includegraphics[width=0.90\linewidth]{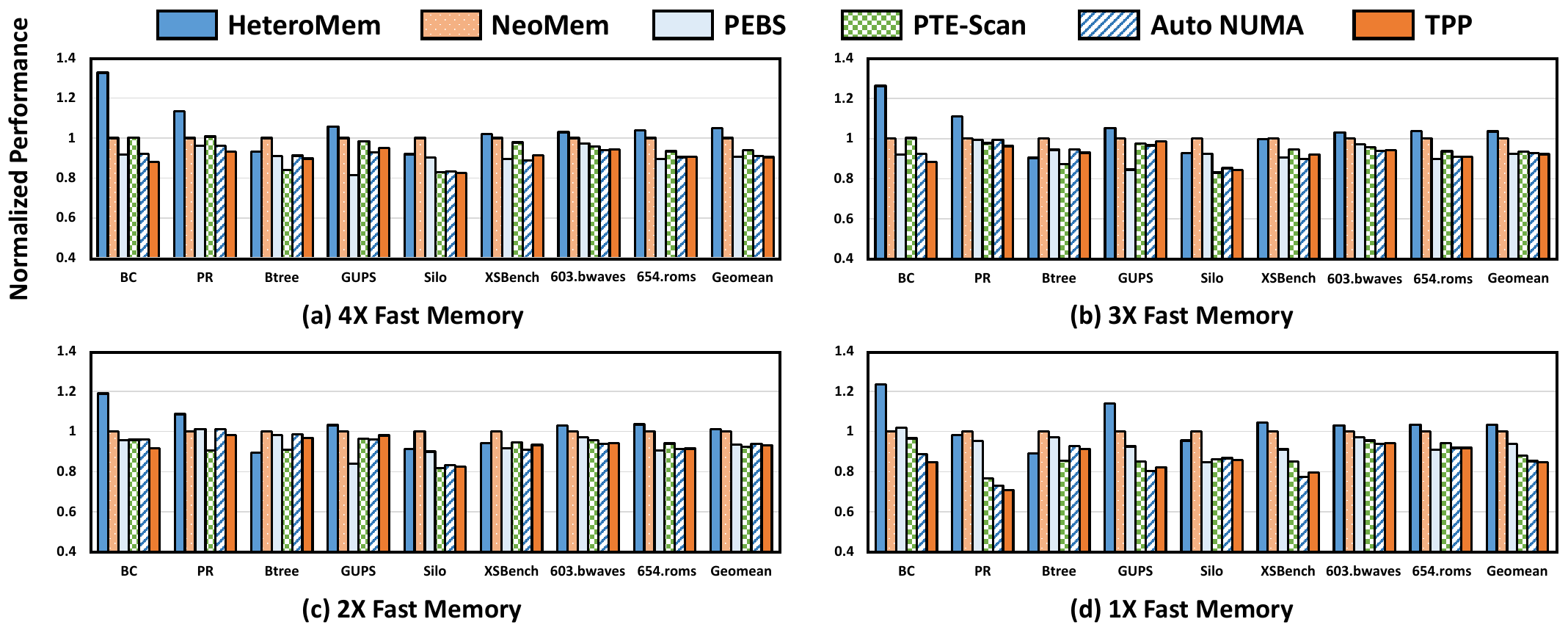} 
    \vspace{-0.2cm}
    \caption{Performance Comparison with Different Fast Memory Size Configuration.}
     \label{fig:performance_comparison}
     \vspace{-0.3cm}
\end{figure*}

Our FPGA prototype of HeteroMem consumes 224.4k ALMs (24.6\%) and 3615 BRAMs (M20K, 27.2\%), with no usage of DSPs. The layout of our HeteroMem FPGA prototype is illustrated in Figure~\ref{fig:implementation}-(a), where the red part represents our HeteroMem logic. This logic acts as an abstraction layer between Intel CXL hard IP and memory controller. Figure~\ref{fig:implementation}-(b) provides a photo of our implementation platform. Finally, Figure~\ref{fig:implementation}-(c) presents an overview of the system emulated by HeteroBox for the subsequent evaluation. 
We configure HeteroBox to emulate two memory regions in CXL-extended memory space: one fast memory region and one slow memory region. Application memory is allocated in CXL-extended memory space, while OS memory remains in the host memory.

\section{Evaluation}
\label{sec:evaluation}

\subsection{Experimental Setup}

\noindent{\textbf{Baselines:}}
We compare the performance of our HeteroMem design with five baseline memory tiering systems: PTE-Scan~\cite{damon}, PEBS~\cite{pebs_events}, Auto NUMA~\cite{corbet2012autonuma}, TPP~\cite{tpp_asplos23}, and NeoMem~\cite{neomem}. PTE-Scan utilizes the access bit in the page table entry for hotness profiling. The access bit is set by the MMU when a page is accessed, and PTE-Scan periodically scans the page table, clearing the access bit and detecting a page as hot if its access bit is set. PTE-Scan migrates hot pages to fast memory as soon as they are detected.
PEBS uses hardware counters to profile page hotness. It is configured to sample the $MEM\_LOAD\_L3\_MISS\_RETIRED$ event. When an LLC miss occurs, the hardware counters increment by one. When the counter reaches a given threshold, it overflows and records the address of the current memory access. The PEBS memory tiering system detects a page as hot and migrates it to fast memory when its address is recorded.
TPP and AutoNUMA utilize hint-fault monitoring to detect hot data in slow memory.
NeoMem is a recently proposed memory tiering system that achieves state-of-the-art performance. It enhances the CXL memory device with a hot page profiling structure, which identifies and reports hot pages to the host CPU for migration to faster memory.

\noindent{\textbf{Benchmarks:}}
We select eight benchmarks for evaluation: GUPS~\cite{gups}, a microbenchmark that performs parallel read-modify-write operations in a uniform or skewed random pattern within its working set; 
Silo~\cite{silo}, an in-memory database engine; 
Btree, an in-memory index lookup; 
XSBench~\cite{xsbench}, an HPC workload; 
Betweenness Centrality (BC) and PageRank (PR)\cite{gapbs}, two graph processing workloads;
and 603.bwaves and 654.roms, two benchmarks from SPEC 2017.
These benchmarks are widely used to evaluate memory tiering systems in previous works\cite{hemem_sosp21, memtis_sosp23, neomem}. 
The Resident Set Size (RSS) of these benchmarks ranges from 4.4GB to 11.1GB. 

\subsection{Validation of HeteroBox}

We use the Intel MLC tool~\cite{mlc} to evaluate the latency and bandwidth attribute of the HeteroBox emulation platform. We first focus on the case of a single region. As illustrated in Figure~\ref{fig:mlc_bandwidth_latency}-(a), we configure the HeteroBox to contain one region and set the latency register value from 0 to 256. Using the Intel MLC, we then measure the region’s latency.
Changing the bandwidth configuration values from 128 to 32, we plot four curves of relationship between the value of latency register and the latency attribute of this region. Additionally, we include a curve called “theoretical latency”, which incorporates the latency from both the CXL link and the CXL IP on the FPGA. 
The theoretical latency also includes the latency introduced by the HeteroBox emulation logic. 
Due to the latency introduced by the underlying memory controller and DDR4 DRAM, the region's latency is the same when the latency register is set to 0 or 32.
All four curves corresponding to different bandwidth configuration values overlap, showing consistent latency performance when compared to theoretical latency.
These results demonstrate two key findings: (1) \textbf{The HeteroBox emulation logic can precisely adjust latency attribute of a region}. (2) \textbf{Modifying the bandwidth configuration register value does not affect the latency attribute of the region}.

As shown in Figure~\ref{fig:mlc_bandwidth_latency}-(b), we plot five curves representing the relationship between the bandwidth configuration register value and the bandwidth attribute of the region. Each curve corresponds to a different latency register value. Additionally, a theoretical bandwidth curve is drawn by using the point where the bandwidth configuration register value is 32 as a base point, and expanding proportionally with the bandwidth configuration register value.
The results demonstrate that: \textbf{The HeteroBox emulation logic can accurately adjust the bandwidth attribute of a region}.



\begin{figure*} [t]
    \centering
    \includegraphics[width=0.85\linewidth]{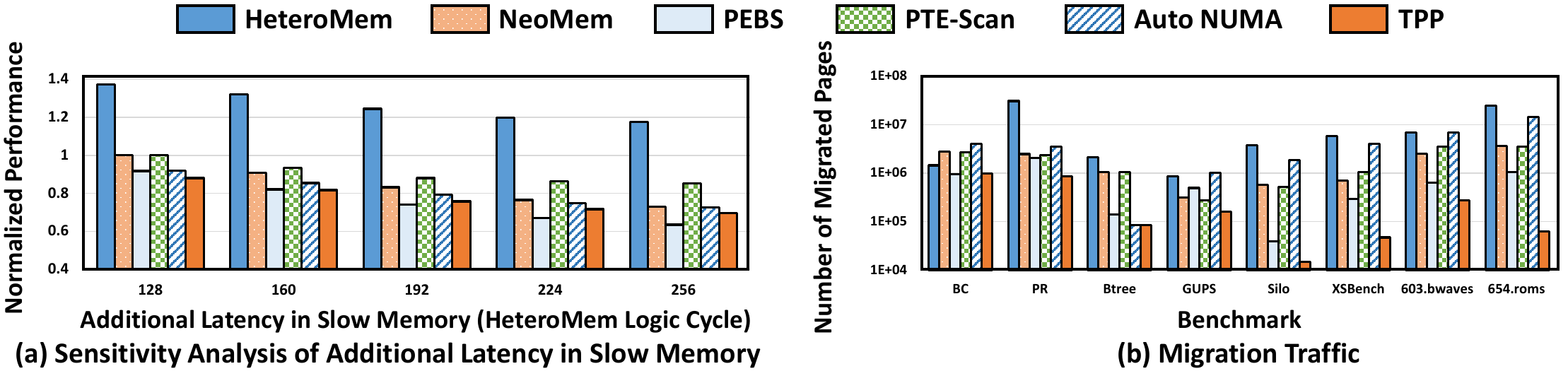} 
    \vspace{-0.4cm}
    \caption{Sensitivity Analysis and Profiling Result of HeteroMem}
         \label{fig:sensitivity_analysis}
         \vspace{-0.3cm}
\end{figure*}

\begin{figure*} [t]
    \centering
    \includegraphics[width=0.85\linewidth]{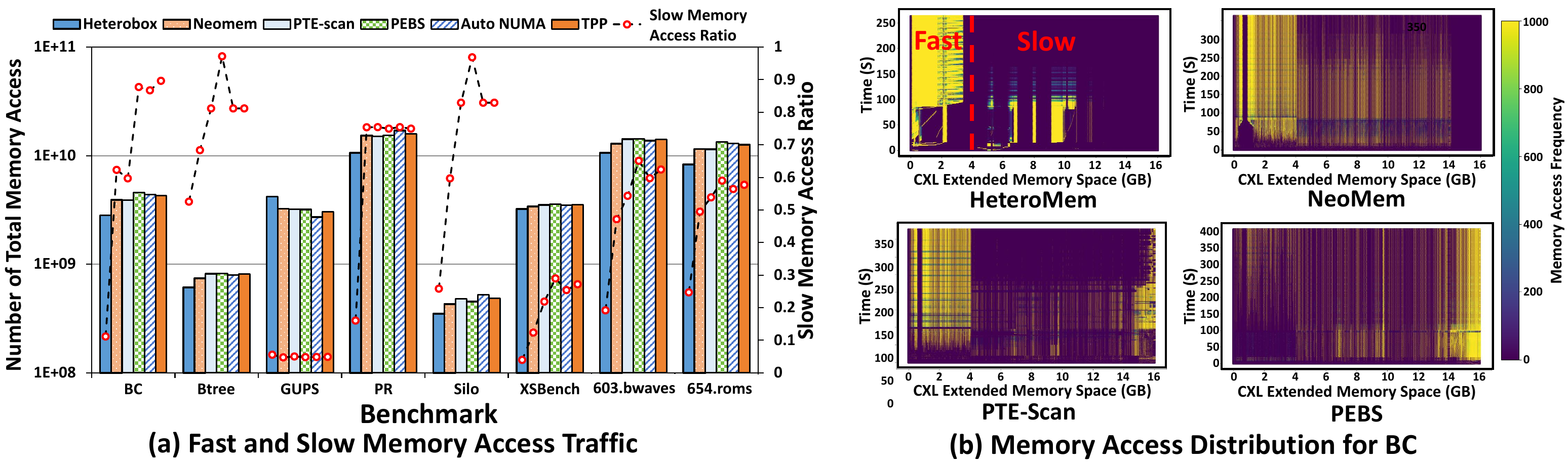} 
    \vspace{-0.4cm}
    \caption{Analysis of Memory Access Distribution.}
     \label{fig:memory_access_distribution}
     \vspace{-0.5cm}
\end{figure*}

Next, we discuss the case of multiple regions. As shown in Figure~\ref{fig:mlc_bandwidth_latency}-(c) and Figure~\ref{fig:mlc_bandwidth_latency}-(d), we configure the HeteroBox with four regions, dividing the FPGA memory into four equal parts. We modify the kernel to create four NUMA nodes corresponding to these regions and use Intel MLC to evaluate their latency and bandwidth attributes.
In Figure~\ref{fig:mlc_bandwidth_latency}-(c), the bandwidth configuration register value for all regions is set to 128, while the latency register values for regions 1, 2, and 3 are set to 32, 64, and 96, respectively. We vary the latency register value of region 4 and plot the latency attributes of all four regions.
In Figure~\ref{fig:mlc_bandwidth_latency}-(d), the latency register value for all regions is set to 128, while the bandwidth configuration register values for regions 1, 2, and 3 are set to 128, 80, and 32, respectively. We vary the bandwidth configuration register value of region 4 and plot the bandwidth attributes of all four regions.
The results show that: (1) \textbf{The HeteroBox emulation logic can accurately adjust the latency and bandwidth attributes in the case of multiple regions}. (2) \textbf{Adjusting the latency or bandwidth of one region does not affect the attributes of the other regions}.

\subsection{Main Results} 
\noindent{\textbf{Performance Comparison:}}
To evaluate the performance of HeteroMem, we configure a 2MB remapping cache and compare HeteroMem's end-to-end performance with baseline systems. We run HeteroMem on HeteroBox emulation platform, dividing the memory space into fast and slow regions. The fast memory region size varies from 4GB to 1GB with no additional latency. The slow memory region has an additional latency of 128 cycles in a 200MHz clock domain. Neither memory region has any bandwidth restrictions.

As shown in Figure~\ref{fig:performance_comparison}, HeteroMem consistently maintains superior geomean performance compared to baseline systems across various fast memory sizes. All performance statistics are normalized to NeoMem. In the 4GB fast memory configuration, HeteroMem's geomean performance is 5.1\% to 16.2\% higher than baseline systems. For BC benchmark, HeteroMem achieves 32.7\% higher performance than NeoMem and PTE-Scan, and 44.6\% higher than PEBS. 
Additionally, the performance improvement of HeteroMem increases with larger fast memory sizes, primarily due to the enhanced effectiveness of HeteroMem's accurate cold page profiling mechanism.
The performance improvement of HeteroMem stems from its accurate hardware-based profiling method and efficient hardware-based data movement mechanism, as described in Section~\ref{sec:heteromem}. 

For the following experiments, we utilize a configuration with 4GB fast memory and a 2MB remapping cache to delve into the superior performance of HeteroMem.


\noindent{\textbf{Different Slow Memory Latency:}}
To explore how slow memory latency affects the performance of HeteroMem, we vary the additional latency in slow memory from 128 cycles to 256 cycles (200MHz clock domain) and test the performance of BC benchmark. The results are shown in Figure~\ref{fig:sensitivity_analysis}-(a). HeteroMem maintains superior performance compared to other memory tiering systems, with non-significant performance degradation as slow memory latency increases. 
This is because HeteroMem accurately and timely migrates hot pages from slow to fast memory.
PEBS performs the worst because it samples hot pages, leading to fewer detected hot pages and less page migration.
TPP also shows low performance due to its hysteretic promotion mechanism.

\noindent{\textbf{Number of Migrated Pages:}}
As shown in Figure~\ref{fig:sensitivity_analysis}-(b), we profile the number of migrated pages for HeteroMem and five baseline systems. For the BC benchmark, all hot pages can fit in fast memory, so HeteroMem stops migrating once they are placed there. NeoMem, PTE-Scan and Auto NUMA show redundant migrations, while PEBS and TPP fail to migrate all hot pages due to slow convergence.
In other benchmarks, where hot pages cannot all fit in fast memory, HeteroMem migrates more pages than the baseline systems. With a 256MB/s migration limit and higher bandwidth, HeteroMem's additional migrations have negligible overhead. This results in a higher proportion of fast memory accesses, leading to improved end-to-end performance of HeteroMem.



\noindent{\textbf{Memory Access Distribution:}}
As shown in Figure~\ref{fig:memory_access_distribution}-(a), we profile the number of memory accesses to fast and slow memory for all six memory tiering systems. 
We show the total number of memory accesses and the ratio of memory accesses to slow memory.
For all benchmarks except GUPS, HeteroMem shows fewer total memory accesses compared to the baseline systems due to the additional CPU memory accesses required by software-based data migration in the baseline systems. In the GUPS benchmark, HeteroMem exhibits more memory accesses because the performance metric of GUPS is based on the memory access count within a given period.
Additionally, HeteroMem demonstrates a lower ratio of slow memory accesses across all benchmarks (except for GUPS) compared to the baseline systems. 
The higher proportion of fast memory accesses explains HeteroMem's superior  performance, showing its effective profiling of hot and cold data and optimal placement of hot data in fast memory.

In Figure~\ref{fig:memory_access_distribution}-(b), we plot the memory access distribution for the BC benchmark. We implement a 32-bit hardware counter for each 2MB physical page in the CXL extended memory and record the memory access count to each 2MB page every second. The results show that, compared to the baseline systems, HeteroMem quickly and accurately migrates all hot data to fast memory, demonstrating the effectiveness of its memory management strategy.

\section{Related Work}
\label{sec:related_work}

\subsection{Software-based memory tiering system}


Many previous works using software-based methods have tried to optimize the performance of memory tiering systems in various aspects, including memory access profiling~\cite{thermostat_asplos17, dancing_ipdps21, re_fault_icbdsc22, damon}, page migration~\cite{page_migration_support_for_disaggregated_non_volatile_memories_memsys19, granularity_aware_page_migration_ics18, nimble_asplos19}, page classification~\cite{ebm, multi-clock}, and huge page strategies~\cite{amp, memtis_sosp23}.
TPP~\cite{tpp_asplos23} proposes to promote a page after it is scanned with the access bit set twice. MEMTIS~\cite{memtis_sosp23} uses histogram-based hot set classification to achieve more stable hot page detection. NOMAD~\cite{nomad_osdi24} employs transactional page migration to mitigate the overhead of page migration.
Additionally, there are works that utilize application-level information to manage data placement at the object level~\cite{memkind, pmdk, hildebrand2020autotm, li2022GCMove, ren2021sentinel, Wang2019Panthera, wei2015_2pp}. However, although these works strive to optimize the performance of memory tiering systems, their performance is limited by the inherent inaccuracy of software profiling and the high overhead of data migration using the CPU, as described in Section~\ref{sec:background}.
Our HeteroMem design employs a hardware-based high-resolution profiling scheme for hot and cold page detection. Meanwhile, HeteroMem uses hardware-based data movement, which achieves high bandwidth and low overhead without wasting CPU cycles and CXL link bandwidth.

\subsection{Hardware-based memory tiering system}


There are also many works that try to optimize memory tiering systems using hardware-based methods. They manage fast memory as a memory cache~\cite{optane_memory_mode} or use fast memory in flat memory mode~\cite{cameo_micro14, slic_fm_hpca17, chameleon_micro18, sehmm_sc10, mempod_hpca17, hopp}. Intel Optane persistent memory provides memory mode~\cite{optane_memory_mode}, allowing users to use DRAM as a cache for non-volatile memory.
CAMEO~\cite{cameo_micro14} combines memory cache mode and flat address mode. It manages fast memory in a cache-like manner, constrains data movement within associated sets, and moves data at cacheline granularity. MemPod~\cite{mempod_hpca17} groups HBM and DRAM connected to CPU into clusters called \emph{Pods} and manages data placement within each Pod.
However, these hardware-based methods, proposed before the CXL era, involve intrusive modifications to CPU, making them inflexible and unable to adapt to dynamic and diverse CXL-based device-side heterogeneous memory systems. 
HeteroMem, on the other hand, places all its hardware logic on the device side and can be runtime configured to adapt to various configurations, offering a one-for-all solution.


A recent work, NeoMem, introduces a method to profile hot data on the device side, achieving state-of-the-art performance. However, NeoMem relies on CPU-side profiling for cold data and uses CPU to manage data movement, while HeteroMem employs device-side hardware logic for profiling both hot and cold data, as well as managing data movement. The design of HeteroMem creates a homogeneous abstraction for the CPU and improves overall performance compared to NeoMem.

\section{Discussion \& Future Work}
\label{sec:discussion}
\noindent{\textbf{Support for CXL memory pooling and sharing:}}
Memory pooling and sharing are promising use cases of CXL~\cite{ha2023dynamic}, allowing memory to be shared between processors~\cite{cxl_pooling_sharing}. 
Memory in a memory pool can be dynamically allocated and deallocated, leading to more efficient use of memory resources. 
Memory sharing, on the other hand, enables each processor to read from or write to the shared memory, facilitating easy data sharing and communication between processors.
The data migration process of HeteroMem is completely transparent to CPU.
As a result, HeteroMem serves as a plug-in solution for both memory pooling and sharing scenarios.



\noindent{\textbf{Hardware overhead and Scalability:}}  
The hardware overhead of HeteroMem remains stable as the CXL-extended memory system scales. Firstly, HeteroMem's hardware overhead increases minimally with the number of CXL devices. HeteroMem can be integrated into a CXL switch, and the number of CXL switches is significantly smaller than the number of CXL devices. Furthermore, not every CXL switch needs to be equipped with HeteroMem hardware. HeteroMem can manage the entire memory space it can access, so it only needs to be implemented in the CXL switch connected to the host CPU.
Secondly, as the memory space scales, the increase in HeteroMem's hardware overhead remains minimal. The hot page profiling component of HeteroMem utilizes Count-Min Sketch to detect hot pages, with an error bound related to the memory access sequence length rather than the memory space size. The cold page profiling component uses two bits of SRAM for every page. This structure only needs to be allocated for fast memory regions with demotion targets, which typically have a small capacity.

\noindent{\textbf{Profiling and Migration Granularity:}}
In our current HeteroMem design, we use a 4KB page for both profiling and migration granularity for design simplicity. Generally speaking, smaller granularity enables more flexible data placement but introduces heavier remapping overhead. Using larger granularity reduces the overhead of remapping but introduces amplification in data movement and wastage of fast memory due to hotness skew in a large page.
There are no theoretical difficulties in implementing profiling and migration granularities other than 4KB or using multiple granularities in HeteroMem. We leave this for our future work.

\section{Conclusion}
\label{sec:conclusion}


In this paper, we introduce HeteroBox, an innovative emulation platform built on a CXL-enabled FPGA. HeteroBox supports emulating a CXL-extended heterogeneous memory system with multiple regions, each with configurable latency and bandwidth characteristics. Based on the HeteroBox emulation platform, we propose HeteroMem, a novel hardware-based, device-side memory tiering system. HeteroMem is transparent to the host CPU and can be runtime configured to adapt dynamically to different CXL-extended heterogeneous memory system configurations. Additionally, HeteroMem achieves high performance through accurate memory access profiling and efficient data movement.
Our evaluation results show that HeteroMem achieves a 5.1\% to 16.2\% performance improvement over existing memory tiering solutions.

\section*{Acknowledgments}

We thank all the reviewers for their valuable comments. This work is supported by National Key Research and Development Program of China (Grant No.2023YFB4502702). This work is also supported by Beijing Natural Science Foundation (Grant No. L243001), National Natural Science Foundation of China (Grant No. 62032001), 111 Project (B18001) and the Natural Science Foundation of China (Grant No. 62332021 and 62472007).

\bibliographystyle{IEEEtranS}
\bibliography{refs}

\end{document}